\documentstyle[myart,12pt]{article}
\normalbaselines
\font\tenbm=cmmib10
\font\sevenbm=cmmib7
\font\fivebm=cmmib5
\newfam\bmfam
\textfont\bmfam=\tenbm \scriptfont\bmfam=\sevenbm
\scriptscriptfont\bmfam=\fivebm
{\count0=\number\bmfam \multiply\count0 by "100
\def\defbgreek#1#2#3{{\count1=\count0 \advance\count1 by "#2#3
  \global\mathchardef#1=\count1 }}
\defbgreek\balpha  0B \defbgreek\brho       1A
\defbgreek\bbeta   0C \defbgreek\bsigma     1B
\defbgreek\bgamma  0D \defbgreek\btau       1C
\defbgreek\bdelta  0E \defbgreek\bupsilon   1D
\defbgreek\bepsilon0F \defbgreek\bphi       1E
\defbgreek\bzeta   10 \defbgreek\bchi       1F
\defbgreek\bmeta   11 \defbgreek\bpsi       20
\defbgreek\btheta  12 \defbgreek\bomega     21
\defbgreek\biota   13 \defbgreek\bvarepsilon22
\defbgreek\bkappa  14 \defbgreek\bvartheta  23
\defbgreek\blambda 15 \defbgreek\bvarpi     24
\defbgreek\bmu     16 \defbgreek\bvarrho    25
\defbgreek\bnu     17 \defbgreek\bvarsigma  26
\defbgreek\bxi     18 \defbgreek\bvarphi    27
\defbgreek\bpi     19}
\input{tcilatex}

\begin{document}

\title{Dynamics of stochastic systems and peculiarities of measurements in them}
\author{Yuri A.Rylov}
\date{Institute for Problems in Mechanics, Russian Academy of Sciences,\\
101, bild.1 Vernadskii Ave., Moscow, 119526, Russia. \\
e-mail: rylov@ipmnet.ru}
\maketitle

\begin{abstract}
Technique of stochastic systems dynamics is constructed. Replacement of the
dynamic system parameters by their effective (averaged) values in the action
for the statistical ensemble of dynamic systems leads to the action for the
ensemble of stochastic systems. Character of stochasticity is determined by
the form of effective parameters. Peculiarities of measurement connected
with simultaneous existence of individual dynamic system and statistical
average system are considered.
\end{abstract}

PACS 2003:\ \ 05.10.Gg, 03.65.Ta, 05.20.Gg


\section{Introduction}

Let us consider some physical system ${\cal S}$. Its time evolution may be
deterministic or stochastic. In the first case the physical system ${\cal S}$
associates with some dynamic system ${\cal S}_{{\rm d}}$, (or deterministic
physical system). Dynamic system ${\cal S}_{{\rm d}}$ is a totality of state
described by dynamic variables $X$ and dynamic equations, describing time
evolution of the state $X$. If number of dynamic variables is finite or
countable, the dynamic system ${\cal S}_{{\rm d}}$ is called discrete
dynamic system. Evolution of the state $X=\left\{ {\bf x},{\bf \dot{x}}%
\right\} $, ${\bf x}=\left\{ x^{\alpha }\right\} ,\;\;\alpha =1,2,...,n$ of
the discrete dynamic system ${\cal S}_{{\rm d}}$ is described by the action
functional 
\begin{equation}
{\cal S}_{{\rm d}}:\;\;{\cal A}\left[ {\bf x}\right] =\int L\left( t,{\bf x},%
{\bf \dot{x},}\right) dt,\qquad {\bf \dot{x}\equiv }\frac{d{\bf x}}{dt}
\label{f1.0}
\end{equation}
Solutions of dynamic equations are extremals of the functional (\ref{f1.0}).
Dynamic equations are obtained by variation of the action with respect to $%
{\bf x}${\bf . } 
\begin{equation}
\frac{\delta {\cal A}}{\delta x^{\alpha }}=\frac{\partial L}{\partial
x^{\alpha }}-\frac{d}{dt}\frac{\partial L}{\partial \dot{x}^{\alpha }}%
=0,\qquad \alpha =1,2,...,n  \label{f1.01}
\end{equation}
The canonical momentum $p=\left\{ p_{0},{\bf p}\right\} =\left\{
p_{i}\right\} ,$ $\left\{ p_{i}\right\} ,\;\;i=1,2,...n$ is associated with
the discrete dynamic system ${\cal S}_{{\rm d}}$%
\begin{equation}
p_{\alpha }=\frac{\partial L}{\partial x^{\alpha }},\qquad \alpha
=1,2,...n;\qquad p_{0}=\frac{\partial L}{\partial \dot{x}^{\beta }}\dot{x}%
^{\beta }-L  \label{f1.02}
\end{equation}

If dynamic variables $X$ describing the state of the dynamic system ${\cal S}%
_{{\rm d}}$ form a continuous set, ${\cal S}_{{\rm d}}$ is a continuous
dynamic system. Dynamic variables $X=u=\left\{ u_{B}\right\} ,\;\;B=1,2,...s$
of continuous dynamic system are labelled by some quantities ${\bf x=}%
\left\{ x^{1},x^{2},...,x^{n}\right\} $. Dependent variables $u=\left\{
u_{B}\right\} ,\;\;B=1,2,...s$ are considered to be functions of independent
variables $x=\left\{ t,{\bf x}\right\} =\left\{ x^{k}\right\}
,\;\;\;k=0,1,...n$, where $t=x^{0}$ is the time.

Continuous dynamic system is described by the action functional 
\begin{equation}
{\cal S}_{{\rm d}}:\;\;{\cal A}\left[ u\right] =\int {\cal L}\left(
u,\partial _{k}u_{B}\right) d^{n+1}x,\qquad \partial _{k}u_{B}\equiv \frac{%
\partial u_{B}}{\partial x^{k}},\qquad d^{n+1}x{\bf =}\prod_{k=0}^{k=n}dx^{k}
\label{f1.1}
\end{equation}
The quantity ${\cal L}$ is called the Lagrange function density.

Dynamic equations for the dynamic system ${\cal S}_{{\rm d}}$ are obtained
as equations for extremals of the action functional (\ref{f1.1}). They are
obtained by means of variation with respect to dependent variables $u$ 
\begin{equation}
\frac{\delta {\cal A}}{\delta u_{B}}=\frac{\partial {\cal L}}{\partial u_{B}}%
-\partial _{k}\left( \frac{\partial {\cal L}}{\partial u_{B,k}}\right)
=0,\qquad B=1,2,...s  \label{f1.4}
\end{equation}
Here and further summation over repeated Latin indices from $0$ to $n$ is
assumed. The flux $j^{k}\left( x\right) $ 
\begin{equation}
j^{k}\left( x\right) =\sum_{B=1}^{B=s}\frac{\partial {\cal L}}{\partial
u_{B,k}}u_{B}  \label{f1.2}
\end{equation}
and the energy-momentum tensor $T_{l}^{k}\left( x\right) $ 
\begin{equation}
T_{l}^{k}\left( x\right) =\sum_{B=1}^{B=s}\frac{\partial {\cal L}}{\partial
u_{B,k}}u_{B,l}-\delta _{l}^{k}{\cal L}  \label{f1.3}
\end{equation}
are physical quantities associated with the continuous dynamic system ${\cal %
S}_{{\rm d}}$.

Let ${\cal S}_{{\rm st}}$ be a discrete stochastic system, i.e. such a
system, whose state $X$ is described by finite or countable number of
variables, but there are no dynamic equation for ${\cal S}_{{\rm st}}$.
Concept of the state for stochastic system ${\cal S}_{{\rm st}}$ is rather
conditional, because there are no dynamic equations, and one can change the
number of variables $X$, describing the state of ${\cal S}_{{\rm st}}$. We
shall consider discrete stochastic system ${\cal S}_{{\rm st}}$ as a
discrete dynamic system ${\cal S}_{{\rm d}}$, whose evolution is spoiled by
influence of some stochastic agent. Then dynamic variables $X$, describing
the state of stochastic system ${\cal S}_{{\rm st}}$ may be considered to be
variables describing the state of corresponding dynamic system ${\cal S}_{%
{\rm d}}$.

Absence of dynamic equations for ${\cal S}_{{\rm st}}$ does not mean that
the evolution of ${\cal S}_{{\rm st}}$ is completely random and has nothing
regular. One can separate regular component ${\cal C}_{{\rm reg}}$ of the
system ${\cal S}_{{\rm st}}$ evolution, if to keep track evolution of many
similar independent systems ${\cal S}_{{\rm st}}$, starting from the
similarly preparing states. Such a set of dynamic or stochastic systems $%
{\cal {S}}$ is called the statistical ensemble ${\cal E}[N,{\cal S}]$ of
systems ${\cal {S}}$, where $N$ is the number of systems ${\cal {S}}$ in the
ensemble ${\cal E}[N,{\cal S}]$.

Let us consider the statistical ensemble ${\cal E}[N,{\cal S}_{{\rm st}}]$,
consisting of many ($N\rightarrow \infty $) stochastic systems ${\cal S}_{%
{\rm st}}$. Because of independence of ${\cal S}_{{\rm st}}$ the random
components of the state evolution are compensated, but regular ones are
accumulated. As a result at $N\rightarrow \infty $ only regular component of
the state evolution remains. The state of the statistical ensemble evolves
regularly (deterministically). Formally this circumstance is displayed in
the fact that the statistical ensemble ${\cal E}[\infty ,{\cal S}_{{\rm st}%
}] $ appears to be a dynamic system, for which there exist dynamic
equations. But ${\cal E}[\infty ,{\cal S}_{{\rm st}}]$ is a continuous
dynamic system, although ${\cal S}_{{\rm st}}$ is a discrete one.

Existence of regular component in evolution of stochastic systems admits one
to work with stochastic systems as with dynamic systems whose evolution is
spoiled by presence of random component. But we must change our approach to
description of the discrete physical system ${\cal S}$ evolution. Any
discrete physical system (${\cal S}_{{\rm st}}$ and ${\cal S}_{{\rm d}}$) is
considered to be a stochastic system ${\cal S}_{{\rm st}}$, which contains
regular evolution component ${\cal C}_{{\rm reg}}$ and stochastic evolution
component ${\cal C}_{{\rm st}}$. Dynamic system ${\cal S}_{{\rm d}}$
(deterministic physical system) is a special case of ${\cal S}$, when the
stochastic evolution component vanishes ${\cal C}_{{\rm st}}=0$. The regular
evolution component ${\cal C}_{{\rm reg}}$ of the discrete physical system $%
{\cal S}$ is described as the statistical average dynamic system $%
\left\langle {\cal S}\right\rangle $. The statistical average system $%
\left\langle {\cal S}\right\rangle $ is a continuous dynamic system even in
the case, when ${\cal S}$ is a discrete dynamic system ${\cal S}_{{\rm d}}$.
In other words, $\left\langle {\cal S}_{{\rm d}}\right\rangle $ does not
coincide with ${\cal S}_{{\rm d}}$, because $\left\langle {\cal S}_{{\rm d}%
}\right\rangle $ is a continuous dynamic system, whereas ${\cal S}_{{\rm d}}$
is a discrete one.

Such a situation arises, because description of ${\cal S}_{{\rm d}}$ as a
partial case of physical system ${\cal S}$ is more informative, than
description of ${\cal S}_{{\rm d}}$ as a discrete dynamic system. The
statistical average system $\left\langle {\cal S}\right\rangle $ describes
both evolution components ${\cal C}_{{\rm reg}}$ and ${\cal C}_{{\rm st}}$.
The regular component ${\cal C}_{{\rm reg}}$ is described, when $%
\left\langle {\cal S}\right\rangle $ is considered to be a continuous
dynamic system. The stochastic component ${\cal C}_{{\rm st}}$ is described
by means of reduction of the action ${\cal A}_{\left\langle {\cal S}%
\right\rangle }$ for the system $\left\langle {\cal S}\right\rangle $ to the
form of the action ${\cal A}_{{\cal S}_{{\rm red}}}$ for the continuous set $%
{\cal S}_{{\rm red}}\left[ {\cal S}_{{\rm d}}\right] $ of identical discrete
dynamic systems ${\cal S}_{{\rm d}}$, interacting between themselves. Then
the form of interaction of ${\cal S}_{{\rm d}}$ labels and describes
implicitly the stochastic evolution component ${\cal C}_{{\rm st}}$.

Formally the stochastic evolution component ${\cal C}_{{\rm st}}$ is
described by introduction of a new dependent dynamic variable. In the
simplest case, when the dynamic system ${\cal S}_{{\rm d}}$ is a free
particle, this additional variable is the velocity ${\bf u}_{{\rm st}}\left(
t,{\bf x}\right) $ describing mean value of stochastic evolution component $%
{\cal C}_{{\rm st}}$ (random velocity). It looks as follows.

Let ${\cal S}_{{\rm st}}$ be stochastic particle, whose state $X$ is
described by variables $\left\{ {\bf x},\frac{d{\bf x}}{dt}\right\} $, where 
${\bf x}$ is the particle position. Evolution of the particle state is
stochastic, and there exist no dynamic equations for ${\cal S}_{{\rm st}}$.
Evolution of the state of ${\cal S}_{{\rm st}}$ contains both regular and
stochastic components. To separate the regular evolution components, we
consider a set (statistical ensemble) ${\cal E}\left[ {\cal S}_{{\rm st}}%
\right] $ of many independent identical stochastic particles ${\cal S}_{{\rm %
st}}$. All stochastic particles ${\cal S}_{{\rm st}}$ start from the same
initial state. It means that all ${\cal S}_{{\rm st}}$ are prepared in the
same way. If the number $N$ of ${\cal S}_{{\rm st}}$ is very large, the
stochastic elements of evolution compensate each other, but regular ones are
accumulated. In the limit $N\rightarrow \infty $ the statistical ensemble $%
{\cal E}\left[ {\cal S}_{{\rm st}}\right] $ turns to a dynamic system, whose
state evolves according to some dynamic equations.

Let the statistical ensemble ${\cal E}_{{\rm d}}\left[ {\cal S}_{{\rm d}}%
\right] $ of deterministic classical particles ${\cal S}_{{\rm d}}$ be
described by the action ${\cal A}_{{\cal E}_{{\rm d}}\left[ {\cal S}_{{\rm d}%
}\left( P\right) \right] }$, where $P$ are parameters describing ${\cal S}_{%
{\rm d}}$ (for instance, mass, charge). Let under influence of some
stochastic agent the deterministic particle ${\cal S}_{{\rm d}}$ turn to a
stochastic particle ${\cal S}_{{\rm st}}$. The action ${\cal A}_{{\cal E}_{%
{\rm st}}\left[ {\cal S}_{{\rm st}}\right] }$ for the statistical ensemble $%
{\cal E}_{{\rm st}}\left[ {\cal S}_{{\rm st}}\right] $ is reduced to the
action ${\cal A}_{{\cal S}_{{\rm red}}\left[ {\cal S}_{{\rm d}}\right] }=%
{\cal A}_{{\cal E}_{{\rm st}}\left[ {\cal S}_{{\rm st}}\right] }$ for some
set ${\cal S}_{{\rm red}}\left[ {\cal S}_{{\rm d}}\right] $ of identical
interacting deterministic particles ${\cal S}_{{\rm d}}$. The action ${\cal A%
}_{{\cal S}_{{\rm red}}\left[ {\cal S}_{{\rm d}}\right] }$ as a functional
of ${\cal S}_{{\rm d}}$ has the form ${\cal A}_{{\cal E}_{{\rm d}}\left[ 
{\cal S}_{{\rm d}}\left( P_{{\rm eff}}\right) \right] }$, where parameters $%
P_{{\rm eff}}$ are parameters $P$ of the deterministic particle ${\cal S}_{%
{\rm d}}$, averaged over the statistical ensemble, and this averaging
describes interaction of particles ${\cal S}_{{\rm d}}$ in the set ${\cal S}%
_{{\rm red}}\left[ {\cal S}_{{\rm d}}\right] $. It means that 
\begin{equation}
{\cal A}_{{\cal E}_{{\rm st}}\left[ {\cal S}_{{\rm st}}\right] }={\cal A}_{%
{\cal S}_{{\rm red}}\left[ {\cal S}_{{\rm d}}\left( P\right) \right] }={\cal %
A}_{{\cal E}_{{\rm d}}\left[ {\cal S}_{{\rm d}}\left( P_{{\rm eff}}\right) %
\right] }  \label{a0.6a}
\end{equation}
In other words, stochasticity of particles ${\cal S}_{{\rm st}}$ in the
ensemble ${\cal E}_{{\rm st}}\left[ {\cal S}_{{\rm st}}\right] $ is replaced
by interaction of ${\cal S}_{{\rm d}}$ in ${\cal S}_{{\rm red}}\left[ {\cal S%
}_{{\rm d}}\right] $, and this interaction is described by a change 
\begin{equation}
P\rightarrow P_{{\rm eff}}  \label{a0.6b}
\end{equation}
in the action ${\cal A}_{{\cal E}_{{\rm d}}\left[ {\cal S}_{{\rm d}}\left(
P\right) \right] }$.

The action for the ensemble ${\cal E}_{{\rm d}}\left[ {\cal S}_{{\rm d}}%
\right] $ of free deterministic particles ${\cal S}_{{\rm d}}$ has the form 
\begin{equation}
{\cal E}_{{\rm d}}\left[ {\cal S}_{{\rm d}}\right] :\qquad {\cal A}_{{\cal E}%
_{{\rm d}}\left[ {\cal S}_{{\rm d}}\left( m\right) \right] }\left[ {\bf x}%
\right] =\int L\left( {\bf x,}\frac{d{\bf x}}{dt}\right) dtd{\bf \xi },
\label{a0.9}
\end{equation}
where the Lagrangian function is described by the relation 
\begin{equation}
L\left( {\bf x,}\frac{d{\bf x}}{dt}\right) =-mc^{2}+\frac{m}{2}\left( \frac{d%
{\bf x}}{dt}\right) ^{2}  \label{a0.8}
\end{equation}
${\bf x}={\bf x}\left( t,{\bf \xi }\right) $, ${\bf \xi }=\left\{ \xi
_{1},\xi _{2},\xi _{3}\right\} $. Here variables ${\bf \xi }$ label the
particles ${\cal S}_{{\rm d}}$ of the statistical ensemble ${\cal E}_{{\rm d}%
}\left[ {\cal S}_{{\rm d}}\right] $. The action (\ref{a0.9}) describes some
fluid without pressure.

The mass $m$ is the only parameter of a free noncharged particle. If the
stochasticity is the quantum one, the change (\ref{a0.6b}) has the form 
\begin{equation}
m\rightarrow m_{{\rm eff}}=m\left( 1-\frac{{\bf u}^{2}}{2c^{2}}+\frac{\hbar 
}{2mc^{2}}{\bf \nabla u}\right)  \label{a0.9a}
\end{equation}
where ${\bf u}={\bf u}\left( t,{\bf x}\right) $ is the mean value of the
stochastic velocity component. Quantum constant $\hbar $ appears here as a
coupling constant, describing connection between the regular and stochastic
components of particle motion. The velocity ${\bf u}$ is supposed to be
small with respect to $c$. Then we must make the change (\ref{a0.9a}) only
in the first term of (\ref{a0.8}), because the same change in the second
term give the quantity of the order $O\left( c^{-2}\right) $. After change (%
\ref{a0.9a}) the action (\ref{a0.9}) turns to the action

\begin{equation}
{\cal E}_{{\rm st}}\left[ {\cal S}_{{\rm st}}\right] :\qquad {\cal A}_{{\cal %
E}_{{\rm d}}\left[ {\cal S}_{{\rm d}}\left( m_{{\rm eff}}\right) \right] }%
\left[ {\bf x,u}\right] =\int L\left( {\bf x,}\frac{d{\bf x}}{dt}\right) +L_{%
{\rm st}}\left( {\bf u},{\bf \nabla u}\right) dtd{\bf \xi },  \label{a0.10}
\end{equation}
\begin{equation}
L_{{\rm st}}\left( {\bf u},{\bf \nabla u}\right) =\frac{m}{2}{\bf u}^{2}-%
\frac{\hbar }{2}{\bf \nabla u}  \label{a0.11}
\end{equation}
where ${\bf x}={\bf x}\left( t,{\bf \xi }\right) $, ${\bf \xi }=\left\{ \xi
_{1},\xi _{2},\xi _{3}\right\} $, but ${\bf u}={\bf u}\left( t,{\bf x}%
\right) $ is a function of $t,{\bf x}$. The action (\ref{a0.10}) also
describes a fluid, but now it is a fluid with a pressure, and its
irrotational flow is described by the Schr\"{o}dinger equation \cite{R002}.

Thus, essentially the change (\ref{a0.9a}) describes {\it procedure of
quantization}, because it introduces quantum constant $\hbar $ and
additional dynamic variable ${\bf u}={\bf u}\left( t,{\bf x}\right) $,
describing stochastic component. The action (\ref{a0.10}) describes the free
quantum particle $\left\langle {\cal S}_{{\rm st}}\right\rangle $
completely, and other additional suppositions (for instance, quantum
principles) are not necessary. In principle, the change (\ref{a0.6b}) admits
some modification or correction of the quantization procedure, provided the
change (\ref{a0.9a}) is replaced by some other proper change. In
conventional quantum theory such a modification of quantization procedure is
impossible, because it is determined by the quantum principles. In other
words, quantum system is only one of possible stochastic systems.

Description of physical system ${\cal S}$ by means of the statistical
average system is a statistical description of ${\cal S}$, because it refers
to many identical systems ${\cal S}$. The statistical description cannot
predict result of a measurement of some physical quantity ${\cal R}$.
Statistical description can predict only probability $w\left( R^{\prime
}\right) $ of measurement result $R^{\prime }$, but not the result $%
R^{\prime }$ itself. It means that any prediction of the statistical
description is tested by a mass experiment ($M$-measurement), i.e. by a set
of many similar single experiments. It is valid even in the case, when
result of prediction is quite definite, for instance, if $w\left( R^{\prime
}\right) =1$, or $w\left( R^{\prime }\right) =0$. To verify that the
prediction $w\left( R^{\prime }\right) =1$ is valid, it is not sufficient to
produce one measurement an to obtain the result $R^{\prime }$. It is
necessary to produce many measurements and to obtain the result $R^{\prime }$
in all cases.

Description of ${\cal S}_{{\rm d}}$ as a discrete dynamic system supposes a
priori knowledge, that ${\cal S}_{{\rm d}}$ is a deterministic physical
system. The fact that ${\cal S}_{{\rm d}}$ is a deterministic physical
system is not tested in experiment. It is supposed that repeated experiments
give the same result, and there is no necessity to verify this. This
approach is valid, only if our world is deterministic completely. It cannot
be used, if there are stochastic physical systems, and one cannot state a
priori, whether the given physical system is deterministic or not.

Describing the discrete dynamic system ${\cal S}_{{\rm d}}$ as a discrete
dynamic system, one predicts result of measurement of some quantity ${\cal R}
$, but not its probability. It is possible only, if all possible physical
systems are deterministic, and one is sure a priori, that the considered
physical system is a dynamic system.

Real world is not deterministic completely. There are stochastic physical
systems (for instance, quantum systems), and one should describe
statistically even those physical systems which are supposed to be
deterministic. If the physical system ${\cal S}$, being in reality ${\cal S}%
_{{\rm d}}$, is described as $\left\langle {\cal S}\right\rangle $, then the
deterministic character of ${\cal S}$ can be obtained from the form of the
statistical description in terms of the statistical average system $%
\left\langle {\cal S}\right\rangle $. In other words, one should describe
any physical system ${\cal S}$ as a stochastic system ${\cal S}_{{\rm st}}$,
using the statistical average system $\left\langle {\cal S}\right\rangle $
as a main object of dynamics, which describes both regular and stochastic
evolution components of physical system. Such an approach to the dynamics of
physical systems will be referred to as dynamic conception of statistical
description (DCSD).

Let ${\cal E}\left[ N,{\cal S}\right] $ be a statistical ensemble,
consisting of $N$ independent identical physical systems ${\cal S}$. If $N$
is large enough ($N\rightarrow \infty $), ${\cal E}\left[ N,{\cal S}\right] $
is a continuous dynamic system independently of whether ${\cal S}$ is
stochastic or deterministic. Let ${\cal A}_{{\cal E}\left[ N,{\cal S}\right]
}$ be the action for the statistical ensemble ${\cal E}\left[ N,{\cal S}%
\right] $. As far as the statistical ensemble consists of identical
independent systems, its action ${\cal A}_{{\cal E}\left[ N,{\cal S}\right]
} $ has the property 
\begin{equation}
{\cal A}_{{\cal E}\left[ aN,{\cal S}\right] }=a{\cal A}_{{\cal E}\left[ N,%
{\cal S}\right] },\text{\qquad }a>0,\qquad a=\text{const,\qquad }N,aN\gg 1
\label{c1.1}
\end{equation}

Let us consider statistical ensemble, whose action ${\cal A}_{\left\langle 
{\cal S}\right\rangle }$ has the form 
\begin{equation}
{\cal A}_{\left\langle {\cal S}\right\rangle }=\lim_{N\rightarrow \infty }%
\frac{1}{N}{\cal A}_{{\cal E}\left[ N,{\cal S}\right] }  \label{c1.2}
\end{equation}
Deterministic physical system, whose action has the form (\ref{c1.2}), will
be referred to as statistical average dynamic system $\left\langle {\cal S}%
\right\rangle $, because the action ${\cal A}_{\left\langle {\cal S}%
\right\rangle }$ is the mean action for the statistical ensemble ${\cal E}%
\left[ N,{\cal S}\right] $. The system $\left\langle {\cal S}\right\rangle $
is the statistical ensemble ${\cal E}\left[ N,{\cal S}\right] $, normalized
to one system. According to definition (\ref{c1.2}) and the property (\ref
{c1.1}) the action ${\cal A}_{\left\langle {\cal S}\right\rangle }$ of the
statistical average system $\left\langle {\cal S}\right\rangle $ is
invariant with respect to transformation 
\begin{equation}
N\rightarrow aN,\text{\qquad }a>0,\qquad a=\text{const}  \label{c1.3}
\end{equation}

Formally the statistical average system $\left\langle {\cal S}\right\rangle $
may be considered as a statistical ensemble consisting of one system ${\cal S%
}$. At the same time according to (\ref{c1.2}) this statistical ensemble $%
\left\langle {\cal S}\right\rangle $ is a deterministic physical system,
because there is an action for $\left\langle {\cal S}\right\rangle \ $ and
this action is constructed of the action for the statistical ensemble ${\cal %
E}\left[ N,{\cal S}\right] $ with very large number $N$ of elements ($%
N\rightarrow \infty $).

For non-relativistic deterministic system ${\cal S}_{{\rm d}}$ the state $X$
is a point in the phase space. The state of the ensemble ${\cal E}\left[ 
{\cal S}_{{\rm d}}\right] $ is the density $d\left( X\right) $ of states $X$
in the phase space. The quantity $dN=d(X)dX$ may be regarded as the number
of the ensemble elements, whose state is found inside the interval $(X,X+dX)$%
. The number $d(X)dX$ of states inside the phase volume $dX$ is always
nonnegative. At the proper normalization the quantity $d\left( X\right) $
can be interpreted as a probability density. In this case of
non-relativistic ${\cal S}_{{\rm d}}$ one can calculate the mean value $%
\left\langle f\left( X\right) \right\rangle $ over the statistical ensemble $%
{\cal E}\left[ {\cal S}_{{\rm d}}\right] $ of any function $f$ of the state $%
X$ of ${\cal S}_{{\rm d}}$.

In the general case of relativistic stochastic system ${\cal S}_{{\rm st}}$
one uses a statistical ensemble ${\cal E}[N,{\cal S}_{{\rm st}}]$ for
description of evolution of the stochastic system ${\cal S}_{{\rm st}}$. The
state of the ensemble ${\cal E}[N,{\cal S}]$ is also determined by the
quantity $d(X)$, which describes distribution (density) of states $X$ of
single elements ${\cal {S}}$ of the ensemble ${\cal E}[N,{\cal S}]$. But in
the case of relativistic system ${\cal S}_{{\rm st}}$ is not a density of
points in the phase space. The state $d\left( X\right) $ is rather
complicated tensor quantity. Rank of tensor $d\left( X\right) $ depends on
the number of particles in the physical system ${\cal S}_{{\rm st}}$.

In any case the quantity $d\left( X\right) $ is not a scalar and cannot be
interpreted as a probability density, because it may be negative. It leads
to the fact that the number $dN$ of the ensemble elements with the state
lying in the interval $(X,X+dX)$, may be negative. Possibility of negative
number of the ensemble elements is connected with existence of dynamic
systems (antisystems), evolving in the direction inverse with respect to the
time increase direction. For instance, a particle is described by its world
line ${\cal L}$ in the space-time. It is supposed that the world line has an
orientation (unit vector $n^{i}$, tangent to ${\cal L}$). This vector
describes direction of the system evolution. If this vector is directed
towards the future, ${\cal L}$ describes evolution of a particle. If it is
directed towards the past, ${\cal L}$ describes evolution of an
antiparticle. Thus, particle and antiparticle distinguish in the direction
of time evolution, i.e. in the sign of the temporal component $n^{0}$ of the
orientation vector $n^{i}$.

As far as in the relativistic case one cannot introduce probability density
of the state $X,$ one cannot calculate, in general, the mean value $%
\left\langle f\left( X\right) \right\rangle $. But one can calculate mean
values $\left\langle A\right\rangle $ of additive quantities $A$: number of
particles $n,$ energy $E,$ momentum ${\bf P}$, angular momentum $M^{\alpha
\beta }$ and their densities. It appears to be possible, because the
statistical ensemble ${\cal E}\left[ N,{\cal S}\right] $ is a set of
independent dynamic systems ${\cal S}$, and, if $E$ is the energy of ${\cal S%
}$, then the energy of ${\cal E}\left[ N,{\cal S}\right] $ is a sum of
energies of single systems ${\cal S}$, constituting ${\cal E}\left[ N,{\cal S%
}\right] $. But the mean value $\left\langle f\left( A\right) \right\rangle $
of some function $f$ of the additive quantities $A=n,E,{\bf P},M^{\alpha
\beta }$ cannot be determined, in general. For instance, $\left\langle
E^{2}\right\rangle $ cannot be calculated, in general. Such a description of
the stochastic system dynamics is less informative, than the conventional
probabilistic description. The loss of description informativeness is
connected with the fact that the distribution $d\left( X\right) $ may be
negative and cannot be regarded as the probability density of the state $X$.

There are several reasons, why in the relativistic case the distribution $%
d\left( X\right) $ may be negative . We point only one of them: description
of production and annihilation of particle pairs.

Particle and antiparticle are two different states of one physical object --
world line. The world line, {\it considered to be a physical object} (but
not as a method of description of the particle evolution) will be denoted by
WL. It should keep in mind that consideration of a particle and an
antiparticle as two different states of WL is connected with the
circumstance that consideration of evolution of the physical system in the
reverse direction is necessary, if we want, that the number of physical
systems conserves in the dynamic theory, and in the same time the theory
could describe such processes as production and annihilation of particle
pairs. If a particle and an antiparticle are considered to be two different
physical systems distinguishing in value of some parameter, then in the
process of particle -- antiparticle annihilation the number of physical
systems decreases by two units. It is unclear, how one could describe a
change of number of physical systems in the classical theory. If one assumes
that WL is a real physical object, then section of WL by hyperplane $t=$%
const (SWL) form two kinds of pointlike objects (particle and antiparticle).
Number $dN$ of this pointlike objects (SWL) in the 3-volume $dS_{k}$ is
defined by the relation 
\[
dN=j^{k}dS_{k} 
\]
where $j^{k}$ is the proportionality coefficient, describing density of SWLs
inside the 3-volume $dS_{k}$. The number $dN$ of SWLs may have any sign,
because component $j^{0}$ may have any sign for fixed $dS_{k}$. We assume
that sign of $dN$ is positive for particles and negative for antiparticles.
The term `SWL' is used as a collective concept with respect to concepts
`particle' and `antiparticle'. If for instance, there are a particle and an
antiparticle, the total number of SWLs is equal to zero, and this number
does not change in the process of annihilation. If we connect a separate
dynamic system with any SWL, then to conserve the number of dynamic systems
(SWLs) in the dynamic theory, it is necessary to consider distributions $%
d(X) $ (in the give case $d\left( X\right) =j^{k}\left( x\right) $), where $%
d $ may be any real number.

The fact that the distribution $d\left( X\right) $ describing the state of
the statistical ensemble ${\cal E}[N,{\cal S}_{{\rm st}}]$ may be negative,
presents some problems for interpretation. In particular, $d(X)$ cannot be
regarded as the probability density. For instance, the state, where the
number of physical objects is equal to $1$, cannot be considered as the
state with one particle. It may be the state with two particles and one
antiparticle, or the state with three particles and two antiparticles, etc..
But only such a description of evolution of the statistical ensemble ${\cal E%
}[\infty ,{\cal S}]$, where the number of physical systems does not change,
is possible in classical theory. Describing relativistic stochastic systems,
we are forced to use dynamic conception of statistical description (DCSD),
which cannot use the probability theory for statistical description. DCSD is
more general, but less informative conception of statistical description,
than the conventional probabilistic statistical description.

In the case of non-relativistic deterministic systems ${\cal {S}}$ it is
possible non-relativistic (probabilistic) version of the statistical
ensemble ${\cal E}[\infty ,{\cal S}]$ description. But in the case of
non-relativistic stochastic systems ${\cal {S}}$ the statistical ensemble $%
{\cal E}[\infty ,{\cal S}]$ should be described, using the general method of
DCSD, because the stochastic component of evolution may be relativistic even
in the case, when the regular component of evolution is non-relativistic.

Considering deterministic physical system ${\cal S}_{{\rm d}}$ as a special
case of physical system ${\cal S}$, we must consider and investigate $%
\left\langle {\cal S}_{{\rm d}}\right\rangle $ instead of ${\cal S}_{{\rm d}%
} $. Only then the methods of description of ${\cal S}_{{\rm d}}$ and of $%
{\cal S}_{{\rm st}}$ appear to be similar. Indeed, if ${\cal S}_{{\rm st}}$
is described as $\left\langle {\cal S}_{{\rm st}}\right\rangle $, we need to
describe ${\cal S}$ as $\left\langle {\cal S}_{{\rm st}}\right\rangle $ even
in the case, when stochastic component of evolution of ${\cal S}$ tends to
zero, and ${\cal S}\rightarrow {\cal S}_{{\rm d}}$. In this case description
of ${\cal S}$ tends to description of $\left\langle {\cal S}_{{\rm d}%
}\right\rangle $, but not to description of ${\cal S}_{{\rm d}}$. From
formal point of view it is very reasonable to describe all ${\cal S}$ as $%
\left\langle {\cal S}_{{\rm st}}\right\rangle $ to avoid a jump in
description, when ${\cal S}$ coincide with ${\cal S}_{{\rm d}}$.

Description of ${\cal S}$ as $\left\langle {\cal S}\right\rangle $, even in
the case, when ${\cal S}={\cal S}_{{\rm d}}$ leads to the circumstance that
the basic object of dynamics is always the statistical average system $%
\left\langle {\cal S}\right\rangle $. It means that the dynamic formalism
deals only with statistical average systems $\left\langle {\cal S}%
\right\rangle $.$\ $But measurements under dynamic system ${\cal S}$ deal
with ${\cal S}$ and $\left\langle {\cal S}\right\rangle $. It is very
important one to distinguish between $\left\langle {\cal S}\right\rangle $
and ${\cal S}$ at consideration of experiments with stochastic and
deterministic physical systems, because confusion of $\left\langle {\cal S}%
\right\rangle $ and ${\cal S}$ leads to misunderstandings and paradoxes.

Our conception of physical system dynamics may be qualified as dynamic
conception of statistical description (DCSD). This conception carries out a
statistical description, because its basic object: statistical average
system $\left\langle {\cal S}\right\rangle $ is constructed on the basis of
many identical objects ${\cal S}$. Simultaneously the statistical
description of physical systems is produced in the non-probabilistic dynamic
form, i.e. without a use of the probability theory.

In DCSD the statistical average system $\left\langle {\cal S}\right\rangle $
is considered to be a basic object of dynamics. The recent development of
the space-time geometry \cite{R90,R91,R02} stimulates such an approach.
According to this space-time theory the space-time geometry is a reason of
the fact that the motion of particle appears to be primordially stochastic.
The role of this stochasticity is the more, the less is the particle mass.
It means that the deterministic motion of particles should be explained via
stochastic one, but not vice versa, as one attempted to do earlier, when the
particle motion in the space-time was considered to be primordially
deterministic.

In the second section different methods of the statistical average system $%
\left\langle {\cal S}\right\rangle $ description are considered. In the
third section peculiarities of measurement in stochastic systems are
investigated.

\section{Methods of description of statistical ensembles}

The pure statistical ensemble ${\cal E}_{{\rm p}}[{\cal S}]$ of physical
systems ${\cal S}$ is considered to be a fundamental object of the theory.
In the non-relativistic theory the pure statistical ensemble ${\cal E}_{{\rm %
p}}\left[ {\cal S}_{{\rm d}}\right] $ of deterministic physical systems $%
{\cal S}_{{\rm d}}$ is defined as a statistical ensemble, whose distribution
function $F_{{\rm p}}\left( t,{\bf x},{\bf p}\right) $ in the phase space of
Hamiltonian variables ${\bf x}=\left\{ x^{\alpha }\right\} ,$\ ${\bf p}%
=\left\{ p_{\alpha }\right\} $, $\alpha =1,2,...n$ may be represented in the
form 
\begin{equation}
{\cal E}_{{\rm p}}\left[ {\cal S}_{{\rm d}}\right] :\qquad F_{{\rm p}}\left(
t,{\bf x},{\bf p}\right) =\rho \left( t,{\bf x}\right) \delta \left( {\bf p}-%
{\bf P}\left( t,{\bf x}\right) \right)  \label{a1.2}
\end{equation}
where $\rho \left( t,{\bf x}\right) $ and ${\bf P}\left( t,{\bf x}\right)
=\left\{ P_{\alpha }\left( t,{\bf x}\right) \right\} $, $\alpha =1,2,...n$
are functions of only time $t$ and generalized coordinates ${\bf x}$. In
other words, the pure ensemble ${\cal E}_{{\rm p}}\left[ {\cal S}_{{\rm d}}%
\right] $ is a dynamic system, considered in the configuration space $V_{n}$
of coordinates ${\bf x}$. Relativistic generalization of the pure
statistical ensemble, defined in the form (\ref{a1.2}) is difficult, and for
such a generalization we use another forms of the pure statistical ensemble
representation. There are several different forms.

The system (\ref{a1.2}) is a fluidlike dynamic system in $n$-dimensional
space of $V_{n}$ of coordinate ${\bf x}$. The action for ${\cal E}_{{\rm p}}%
\left[ {\cal S}_{{\rm d}}\right] $ is written as a sum of actions for
independent dynamic systems ${\cal S}_{{\rm d}}$, which are described by the
Lagrangian function $L\left( t,{\bf x},\frac{d{\bf x}}{dt}\right) $. The
action functional for the dynamic system, described by the distribution
function (\ref{a1.2}) can be written as follows 
\begin{equation}
{\cal E}_{{\rm p}}\left[ {\cal S}_{{\rm d}}\right] :\qquad {\cal A}_{L}[{\bf %
x}]=\int\limits_{V_{\xi }}{\rm d}^{n}{\bf \xi }\int L(t,{\bf x},\frac{d{\bf x%
}}{dt}){\rm d}t,\qquad {\rm d}^{n}{\bf \xi }=\prod\limits_{\alpha
=1}^{\alpha =n}{\rm d}\xi _{\alpha },  \label{a2.3}
\end{equation}
where dependent variables ${\bf x}=\{x^{\alpha }\},\;\;\alpha =1,2,...n,\;$%
are considered to be functions of independent variables $\xi =\left\{ t,{\bf %
\xi }\right\} =\left\{ \xi _{0},{\bf \xi }\right\} =\left\{ \xi _{k}\right\}
,\;\;k=0,1,...n$. Variables ${\bf \xi }$ label elements ${\cal S}_{{\rm d}}$
of the statistical ensemble and change inside the region $V_{\xi }$.

The number $N$ of elements ${\cal S}_{{\rm d}}$ in the statistical ensemble $%
{\cal E}_{{\rm p}}\left[ {\cal S}_{{\rm d}}\right] $ is defined by the
relation 
\begin{equation}
N=\int\limits_{V_{\xi }}{\rm d}^{n}{\bf \xi =}\int\limits_{V_{x}}\rho {\rm d}%
^{n}x,\qquad \rho =\frac{\partial \left( \xi _{1},\xi _{2},...\xi
_{n}\right) }{\partial \left( x^{1},x^{2},...x^{n}\right) }  \label{b1.1}
\end{equation}
where $V_{x}$ is the region in the space $V_{n}$ of coordinates ${\bf x}$,
occupied by the fluid. The number $N$ is supposed to be very large $%
N\rightarrow \infty $. Properties of the statistical ensemble do not depend
on the number $N$ of its elements, provided $N$ is large enough. The action (%
\ref{a2.3}) is appropriate for description of relativistic dynamic systems $%
{\cal S}_{{\rm d}}$. Generalization on the case of stochastic systems ${\cal %
S}_{{\rm st}}$ is also possible, although it is not so evident.

Let us make a change of variables in (\ref{a2.3}) and consider independent
variables ${\bf \xi }$ as functions of variables ${\bf x}$, or functions $%
\xi =\left\{ \xi _{i}\right\} ,$ $i=0,1,...n$ as functions of variables $%
x=\left\{ x^{i}\right\} $, $\;i=0,1,...n.$ with $\xi _{0}$ being a
fictitious variable, and $x^{0}=t$ being the time. Then after integration of
some dynamic equations the action (\ref{a2.3}) takes the form \cite{R99,R002}

\begin{equation}
{\cal E}_{{\rm p}}\left[ {\cal S}_{{\rm d}}\right] :\;\;{\cal A}%
_{E}[j,\varphi ,{\bf \xi }]=\int \{L(x^{0},{\bf x},\frac{{\bf j}}{j^{0}}%
)j^{0}-j^{i}p_{i}\}{\rm d}^{n+1}x,  \label{a2.18}
\end{equation}
\begin{equation}
p_{i}=b_{0}\left[ \partial _{i}\varphi +g^{\alpha }({\bf \xi })\partial
_{i}\xi _{\alpha }\right] ,\qquad i=0,1,...,n,,\qquad \partial _{i}\equiv 
\frac{\partial }{\partial x^{i}}  \label{a1.5}
\end{equation}
where dependent variables $\rho ,\;\varphi ,$\ ${\bf \xi }=\left\{ \xi
_{\alpha }\right\} ,$ \ $\alpha =1,2,...n$ of the dynamic system ${\cal E}_{%
{\rm p}}\left[ {\cal S}_{{\rm d}}\right] $ are functions of independent
variables $t,{\bf x}$, where ${\bf x}=\left\{ x^{\alpha }\right\} $, $\alpha
=1,2,...n$. Here and below a summation is produced over repeating Greek
indices from $1$ to $n$ and over Latin ones from $0$ to $n$. The flux of
this fluid is described by the vector $j^{i}$%
\begin{equation}
j^{i}=\frac{\partial J}{\partial \xi _{0,i}}=\frac{\partial \left( x^{i},\xi
_{1},\xi _{2},...\xi _{n}\right) }{\partial \left(
x^{0},x^{1},...x^{n}\right) }=J\frac{\partial x^{i}}{\partial x^{0}},\qquad
i=0,1,...n,\qquad \rho =j^{0}=\frac{\partial J}{\partial \xi _{0,0}}
\label{a2.6}
\end{equation}
where the Jacobian $J$ is considered to be multilinear function of variables 
$\xi _{i,k}\equiv \partial _{k}\xi _{i},\quad i,k=0,1,...n$ 
\begin{equation}
J\equiv \frac{\partial \left( \xi _{0},\xi _{1},...\xi _{n}\right) }{%
\partial \left( x^{0},x^{1},...x^{n}\right) }\equiv \det ||\xi
_{i,k}||,\qquad \xi _{i,k}\equiv \frac{\partial \xi _{i}}{\partial x^{k}}%
\equiv \partial _{k}\xi _{i}\qquad i,k=0,1,...n  \label{a2.5}
\end{equation}
The variable $\rho =j^{0}$ describes density of the ensemble elements
(density of the fluid). The quantity $b_{0}$ is an arbitrary nonvanishing
constant, the functions ${\bf g}\left( {\bf \xi }\right) =\left\{ g^{\alpha
}\left( {\bf \xi }\right) \right\} ,\;\;\alpha =1,2,...n$ are arbitrary
functions of argument ${\bf \xi .}$ Both $b_{0}$ and ${\bf g}$ are a result
of integration of dynamic equations which was made in the process of a
change of variables \cite{R99}.

Attributes of dynamic system ${\cal E}_{{\rm p}}\left[ {\cal S}_{{\rm d}}%
\right] $ are the number $N$ of dynamic systems ${\cal S}_{{\rm d}}$
(particles), flux $j^{k}$ of ${\cal S}_{{\rm d}}$ (particles), the
energy-momentum tensor $T_{l}^{k}$ 
\begin{equation}
T_{l}^{k}=-p_{l}j^{k}-\left\{ L(x^{0},{\bf x},\frac{{\bf j}}{j^{0}}%
)j^{0}-j^{i}p_{i}\right\} \delta _{l}^{k},  \label{c2.1}
\end{equation}
and the angular momentum density $M^{ikl}$. The momentum density $p_{k}$ and
the particle flux $j^{k}$ are expressed by equations (\ref{a1.5}) and (\ref
{a2.6}) respectively. The number $N$ of ${\cal S}_{{\rm d}}$ in ${\cal E}_{%
{\rm p}}\left[ {\cal S}_{{\rm d}}\right] $ is described by the relation (\ref
{b1.1}). Transformation of the number $N$ of the form (\ref{c1.3}) 
\begin{equation}
N\rightarrow \tilde{N}=a^{n}N  \label{c2.1a}
\end{equation}
arises at the transformation of $\xi $ of the form 
\begin{equation}
{\bf \xi }\rightarrow {\bf \tilde{\xi}}=a{\bf \xi }  \label{c2.b}
\end{equation}
At this transformation 
\begin{equation}
j^{k}\rightarrow \tilde{j}^{k}=a^{n}j^{k},\qquad g^{\alpha }({\bf \xi }%
)\rightarrow \tilde{g}^{\alpha }({\bf \tilde{\xi}})=g^{\alpha }({\bf \xi }%
)/a.  \label{c2.2}
\end{equation}
Then the action (\ref{a2.18}) transforms in the same way as $N$, i.e. 
\begin{equation}
{\cal A}\rightarrow a^{n}{\cal A}.  \label{c2.3}
\end{equation}
It means that the action (\ref{a2.18}) describes a statistical ensemble. The
action (\ref{a2.18}) describes statistical ensemble of deterministic
physical systems. It follows from the fact that the action (\ref{a2.18})
contain derivatives only in direction of the vector $j^{k}$ in the form $%
j^{k}\partial _{k}$ \ \ ($\partial _{k}j^{k}=0$). In this case the system of
partial differential equations reduces to a system of ordinary differential
equations. For the action (\ref{a2.18}) to describe a statistical ensemble
of stochastic systems, it is sufficient to modify it in such a way, that the
action stops to depend only on $j^{k}\partial _{k}$, but at transformations (%
\ref{c2.b}), (\ref{c2.2}) the action (\ref{a2.18}) continues to transform
according to (\ref{c2.3}).

Eliminating the variables $j^{k}$ from the action (\ref{a2.18}) \cite{R002},
one obtains instead of (\ref{a2.18}) 
\begin{equation}
{\cal E}_{{\rm p}}\left[ {\cal S}_{{\rm d}}\right] :\qquad {\cal A}_{E}[\rho
,\varphi ,{\bf \xi }]=\int \rho \{-H\left( t,{\bf x,p}\right) -p_{0}\}{\rm d}%
^{n+1}x,  \label{a1.4}
\end{equation}
The quantity $H\left( t,{\bf x},{\bf p}\right) =H\left( x^{0},{\bf x},{\bf p}%
\right) $ is the Hamilton function of the dynamic system ${\cal S}_{{\rm d}}$%
. The variable $\rho =j^{0}$ is defined by (\ref{a2.6}), and the quantities $%
p_{k}$ are defined by relations (\ref{a1.5}). The dynamic variables ${\bf %
\xi }$ are regarded as Lagrangian coordinates, or hydrodynamic potentials
(Clebsch potentials \cite{C57,C59}) which are constant along world line of
any fluid particle.

All three forms (\ref{a2.3}), (\ref{a1.4}), (\ref{a2.18}) of the action are
equivalent and describe the same variational problem \cite{R002}.

By means of a change of variables the action (\ref{a1.4}) can be transformed
to a description in terms of a wave function \cite{R99}. Let us introduce $k$%
-component complex function. $\psi =\{\psi _{\alpha }\},\;\;\alpha
=1,2,\ldots k$, defining it by the relations 
\begin{equation}
\psi _{\alpha }=\sqrt{\rho }e^{i\varphi }u_{\alpha }({\bf \xi }),\qquad \psi
_{\alpha }^{\ast }=\sqrt{\rho }e^{-i\varphi }u_{\alpha }^{\ast }({\bf \xi }%
),\qquad \alpha =1,2,\ldots k  \label{s5.4}
\end{equation}
\[
\psi ^{\ast }\psi \equiv \sum_{\alpha =1}^{k}\psi _{\alpha }^{\ast }\psi
_{\alpha } 
\]
where (*) means the complex conjugate, $u_{\alpha }({\bf \xi })$, $\;\alpha
=1,2,\ldots k$ are functions of only variables ${\bf \xi }$. They satisfy
the relations 
\begin{equation}
-\frac{i}{2}\sum_{\alpha =1}^{k}(u_{\alpha }^{\ast }\frac{\partial u_{\alpha
}}{\partial \xi _{\beta }}-\frac{\partial u_{\alpha }^{\ast }}{\partial \xi
_{\beta }}u_{\alpha })=g^{\beta }({\bf \xi }),\qquad \beta =1,2,...n,\qquad
\sum_{\alpha =1}^{k}u_{\alpha }^{\ast }u_{\alpha }=1  \label{s5.5}
\end{equation}
$k$ is such a natural number that equations (\ref{s5.5}) admit a solution.
In general, $k$ depends on the form of the arbitrary integration functions $%
{\bf g}=\{g^{\beta }({\bf \xi })\}$,\ $\beta =1,2,...n.$

It is easy to verify, that 
\begin{equation}
\rho =\psi ^{\ast }\psi ,\qquad p_{l}=-\frac{ib_{0}}{2\psi ^{\ast }\psi }%
(\psi ^{\ast }\partial _{l}\psi -\partial _{l}\psi ^{\ast }\cdot \psi
),\qquad l=0,1,...n  \label{s5.6}
\end{equation}
The variational problem with the action (\ref{a1.4}) appears to be
equivalent \cite{R99} to the variational problem with the action functional 
\begin{eqnarray}
{\cal A}[\psi ,\psi ^{\ast }] &=&\int \left\{ \frac{ib_{0}}{2}(\psi ^{\ast
}\partial _{0}\psi -\partial _{0}\psi ^{\ast }\cdot \psi )\right.  \nonumber
\\
&&\left. -H\left( x,-\frac{ib_{0}}{2\psi ^{\ast }\psi }(\psi ^{\ast }{\bf %
\nabla }\psi -{\bf \nabla }\psi ^{\ast }\cdot \psi )\right) \psi ^{\ast
}\psi \right\} {\rm d}^{n+1}x  \label{s5.8}
\end{eqnarray}
where ${\bf \nabla }=\left\{ \partial _{\alpha }\right\} ,\;\;\alpha
=1,2,...n$.

At the description by means of the action (\ref{s5.8}) the transformation (%
\ref{c2.1a}) of the number $N$ of ${\cal S}_{{\rm d}}$ arises at the
transformation

\begin{equation}
\psi \rightarrow \tilde{\psi}=\alpha ^{n/2}\psi  \label{c2.4}
\end{equation}
when the flux ${\bf j}$ and density $\rho =j^{0}$%
\begin{equation}
{\bf j}=-\frac{ib_{0}}{2}(\psi ^{\ast }{\bf \nabla }\psi -{\bf \nabla }\psi
^{\ast }\cdot \psi ),\qquad \rho =\psi ^{\ast }\psi  \label{c2.5}
\end{equation}
transforms according to (\ref{c2.2}). Then action (\ref{s5.8}) transforms as
(\ref{c2.3}), and hence it is the action for a statistical ensemble. As in
the case of action (\ref{a2.18}) the action (\ref{a1.4}) and (\ref{s5.8})
also contain derivatives only in combination $j^{k}\partial _{k}$ ($\partial
_{k}j^{k}=0$), but it is not seen explicitly. To show this for (\ref{a1.4}),
one should use definition of the Hamilton function 
\[
H\left( {\bf x},{\bf p}\right) ={\bf \dot{x}p}-L\left( {\bf x,\dot{x}}%
\right) ,\qquad {\bf p}=\frac{\partial L}{\partial {\bf \dot{x}}} 
\]
Then taking into account that ${\bf j}={\bf \dot{x}}j^{0}={\bf \dot{x}}\rho $%
, the Lagrange function density of (\ref{a1.4}) 
\[
\rho \{-H\left( t,{\bf x,p}\right) -p_{0}\}=\rho L\left( {\bf x,}\frac{{\bf j%
}}{j^{0}}\right) -{\bf jp}-\rho p_{0}=\rho L\left( {\bf x,}\frac{{\bf j}}{%
j^{0}}\right) -j^{k}p_{k} 
\]
reduces to Lagrangian density of (\ref{a2.18}). The same situation with the
action (\ref{s5.8}), whose Lagrange function density reduces to Lagrange
function density of (\ref{a1.4}).

For the action (\ref{a2.18}) to describe a statistical ensemble of
stochastic systems, it is sufficient that the action contains derivatives
not only in direction of the flux vector $j^{k}$ and in the same time the
action (\ref{a2.18}) transforms according to (\ref{c2.3}) at the
transformation (\ref{c2.1a}) of the number of the ensemble elements.

Let us imagine now that the Lagrangian function in the action (\ref{a2.18})
depends additionally on the arguments $s_{k}=h\partial _{k}\log \rho
=h_{k}\partial _{k}\log j^{0}$,$\;\;s_{kl}=h_{kl}\partial _{k}\partial
_{l}\log \rho $,\ $k,l=0,1,$ $...n$, (there is no summation over $k$ and $l$
here), where $h_{k}$, $h_{kl}$ are some constants. Arguments $s_{k}$ and $%
s_{kl}$ are invariant with respect to transformation (\ref{c2.4}), (\ref
{c2.1a}). If the Lagrangian function $L$ in (\ref{a2.18}) depends on $s_{k}$
and $s_{kl}$, it transforms under transformation (\ref{c2.1a}) in the same
way as if it would not depend on $s_{k}$ and $s_{kl}$. It means that it
remains to be the action for some statistical ensemble of independent
physical systems. But now the action contains derivatives not only in the
direction of the vector $j^{k}$. Then the system of dynamic equations is
essentially such a system of partial differential equations, which cannot be
reduced to a system of ordinary differential equations. Hence, the
statistical ensemble cannot be considered to be constituted of dynamic
systems, because in this case dynamic equations cannot be reduced to a
system of ordinary differential equations. Such a situation can be
interpreted in the sense that elements of the statistical ensemble are
stochastic systems, and there exist no dynamic equations for them.

Of course, the same consideration is valid for other forms (\ref{a1.4}), (%
\ref{a2.3}), (\ref{s5.8}) of the action. In particular, if the Hamilton
function $H\left( {\bf x},{\bf p}\right) $ in (\ref{s5.8}) is taken in the
form 
\begin{equation}
H_{{\rm eff}}=mc^{2}+\frac{{\bf p}^{2}}{2m}+\frac{\hbar ^{2}}{8m}\left(
\nabla \ln \rho \right) ^{2},  \label{b1.5}
\end{equation}
the dynamic equation for the statistical average system $\left\langle {\cal S%
}\right\rangle $ has the form of the free Schr\"{o}dinger equation, provided
besides {\it the arbitrary constant }$b_{0}${\it \ is chosen to be equal to
quantum constant }$\hbar ${\it , and the case of the irrotational flow of
the fluid }is considered{\it .} This case, when the wave function $\psi $
can have only one component, is considered in \cite{R99,R002}.

It is worth to note that consideration of the statistical ensemble ${\cal E}%
\left[ {\cal S}_{{\rm st}}\right] $ with the effective Hamilton function,
depending on variables $s_{k}$ and $s_{kl}$ is equivalent to consideration
of some set ${\cal S}_{{\rm q}}\left[ {\cal S}_{{\rm d}}\right] $ of
interacting identical dynamic systems ${\cal S}_{{\rm d}},$which is
described by the action (\ref{a0.10}). For instance, when the effective
Hamilton function is written in the form (\ref{b1.5}), the corresponding
action 
\begin{equation}
{\cal S}_{{\rm q}}\left[ {\cal S}_{{\rm d}}\right] :\qquad {\cal A}_{{\rm q}}%
\left[ {\bf x},{\bf u}_{{\rm st}}\right] {\bf =}\int \left\{ \frac{m{\bf 
\dot{x}}^{2}}{2}+\frac{m{\bf u}_{{\rm st}}^{2}}{2}-\frac{\hbar }{2}{\bf %
\nabla u}_{{\rm st}}\right\} dtd^{3}{\bf \xi }  \label{g1.19}
\end{equation}
coincides with the action (\ref{a2.3}), obtained after the change (\ref
{a0.9a}) in the action for the ensemble of deterministic particle.\label{01}
Here ${\bf x}={\bf x}\left( t,{\bf \xi }\right) $, ${\bf u}_{{\rm st}}={\bf u%
}_{{\rm st}}\left( t,{\bf x}\right) $ is a new dynamic variable and ${\bf %
\nabla =}\left\{ \frac{\partial }{\partial x^{1}},\frac{\partial }{\partial
x^{2}},\frac{\partial }{\partial x^{3}}\right\} $, The new dynamic variable $%
{\bf u}_{{\rm st}}$ is a function of ${\bf x}$ and depends on ${\bf \xi }$
via ${\bf x}$. The quantity ${\bf u}_{{\rm st}}$ may be regarded as the mean
velocity of stochastic component. The last term in (\ref{g1.19}) describes
influence of the stochasticity on the regular evolution component. Dynamic
equations in terms of independent variables $\left( t,{\bf x}\right) $,
generated by the action (\ref{g1.19}) for dependent dynamic variables $\rho $%
, ${\bf u=}\frac{d{\bf x}}{dt}${\bf ,} ${\bf u}_{{\rm st}}${\bf ,} ${\bf \xi 
}$ have the form 
\[
\frac{\partial \rho }{\partial t}+{\bf \nabla }\left( \rho {\bf u}\right)
=0,\qquad m\left( \frac{\partial {\bf u}}{\partial t}+\left( {\bf u\nabla }%
\right) {\bf u}\right) =-{\bf \nabla }U_{{\rm B}}\left( t,{\bf x}\right) 
\]
\[
{\bf u}_{{\rm st}}=-\frac{\hbar }{2m}{\bf \nabla }\ln \rho ,\qquad \frac{%
\partial {\bf \xi }}{\partial t}+\left( {\bf u\nabla }\right) {\bf \xi }=0 
\]
where 
\[
\rho =\frac{\partial \left( \xi _{1},\xi _{2},\xi _{3}\right) }{\partial
\left( x^{1},x^{2},x^{3}\right) },\qquad {\bf u}=\frac{\partial \left( {\bf %
x,}\xi _{1},\xi _{2},\xi _{3}\right) }{\partial \left( t,\xi _{1},\xi
_{2},\xi _{3}\right) },\qquad U_{{\rm B}}\left( t,{\bf x}\right) =-\frac{%
\hbar ^{2}}{2m}\frac{1}{\sqrt{\rho }}\nabla ^{2}\sqrt{\rho } 
\]
where $U_{{\rm B}}\left( t,{\bf x}\right) $ is the Bohm potential,
considered to be a function of $t,{\bf x}${\bf .} Introducing 
\[
R=\rho ^{-1}=\frac{\partial \left( x^{1},x^{2},x^{3}\right) }{\partial
\left( \xi _{1},\xi _{2},\xi _{3}\right) } 
\]
as a multilinear function of variables $x^{\alpha ,\beta }=x^{\alpha ,\beta
}\left( t,{\bf \xi }\right) \equiv \partial x^{\alpha }/\partial \xi _{\beta
}$ and taking into account that 
\[
\frac{\partial }{\partial x^{\alpha }}=\frac{\partial \xi _{\beta }}{%
\partial x^{\alpha }}\frac{\partial }{\partial \xi _{\beta }}=\frac{1}{R}%
\frac{\partial R}{\partial x^{\alpha ,\beta }}\frac{\partial }{\partial \xi
_{\beta }}, 
\]
one can write dynamic equations in terms of independent variables $t,{\bf %
\xi }$ 
\begin{equation}
m\ddot{x}^{\alpha }=\frac{\hbar ^{2}}{2mR}\frac{\partial R}{\partial
x^{\alpha ,\beta }}\frac{\partial }{\partial \xi _{\beta }}\left[ \frac{1}{%
\sqrt{R}}\frac{\partial R}{\partial x^{\mu ,\nu }}\frac{\partial }{\partial
\xi _{\nu }}\left( \frac{1}{R}\frac{\partial R}{\partial x^{\mu ,\sigma }}%
\frac{\partial }{\partial \xi _{\sigma }}\frac{1}{\sqrt{R}}\right) \right]
\label{b1.6}
\end{equation}
\[
u_{{\rm st}}^{\alpha }\left( t,{\bf \xi }\right) =\frac{\hbar }{2mR}\frac{%
\partial R}{\partial x^{\alpha ,\beta }}\frac{\partial }{\partial \xi
_{\beta }}\frac{1}{R}, 
\]
This example shows that the free non-relativistic quantum particle is a
stochastic system ${\cal S}_{{\rm st}}$, whose regular component of
evolution is described by the statistical average system $\left\langle {\cal %
S}_{{\rm st}}\right\rangle $. Dynamic equation for $\left\langle {\cal S}_{%
{\rm st}}\right\rangle $ is the free Schr\"{o}dinger equation. Technique of
quantum mechanics deals with $\left\langle {\cal S}_{{\rm st}}\right\rangle $
only and ignores ${\cal S}_{{\rm st}}$ completely. But at consideration of
quantum measurements one deals with both physical systems ${\cal S}_{{\rm st}%
}$ and $\left\langle {\cal S}_{{\rm st}}\right\rangle $. Considering quantum
measurements, one confuses sometimes ${\cal S}_{{\rm st}}$ and $\left\langle 
{\cal S}_{{\rm st}}\right\rangle $. It leads sometimes to misunderstandings
and paradoxes.

Dynamic equations (\ref{b1.6}) may be considered as equations for continuous
set of identical deterministic particles labelled by parameters ${\bf \xi .}$
These particles interact between themselves via some self-consistent
potential $U_{{\rm B}}\left( t,{\bf x}\right) $. Thus, ensemble of
independent stochastic systems ${\cal E}\left[ {\cal S}_{{\rm st}}\right] $
may be considered to be a set ${\cal S}_{{\rm red}}\left[ {\cal S}_{{\rm d}}%
\right] $ of interacting deterministic systems ${\cal S}_{{\rm d}}$, and the
form of interaction describes the character of stochasticity. At
probabilistic description of stochastic systems (for instance, at
conventional description of Brownian particles) the character of
stochasticity is described by means of the probability of transition $%
w\left( X,X^{\prime }\right) $ from the state \ $X$ to the state $X^{\prime
} $. In DCSD a character of stochasticity is described by the form of
interaction between deterministic systems ${\cal S}_{{\rm d}}$ in the
continuous set ${\cal S}_{{\rm red}}\left[ {\cal S}_{{\rm d}}\right] ={\cal E%
}\left[ {\cal S}_{{\rm st}}\right] $ of deterministic systems ${\cal S}_{%
{\rm d}}$.

Generalization of the action to the case of stochastic systems has the
simplest form in the case (\ref{s5.8}), when the action is described in
terms of wave function. For the action to be an action of a statistical
ensemble, it is sufficient that the Lagrange function density would be the
first order uniform function of the combination $\psi ^{\ast }\psi $.

\section{Measurement in stochastic systems}

Let us consider some problems of measurement in stochastic systems ${\cal S}%
_{{\rm st}}$ in the example of quantum mechanics which is a special case of
dynamics of stochastic systems. Quantum mechanics is an axiomatic version of
stochastic system dynamics, where the wave function $\psi $ is considered to
be a fundamental object of dynamics, whereas in DCSD the wave function is
only a method of description of statistical ensemble. In quantum mechanics
the Lagrange function density for any quantum system ${\cal S}_{{\rm q}}$ is
supposed to be a linear function of $\psi $ and linear function of $\psi
^{\ast }$. Then at the transformation (\ref{c2.1}) the condition (\ref{c2.3}
) is fulfilled, and the considered dynamic system ${\cal S}_{{\rm q}}$,
described by means of the wave function, may be considered to be a pure
statistical ensemble ${\cal S}_{{\rm q}}={\cal E}_{{\rm p}}\left[ {\cal S}_{%
{\rm st}}\right] $. If besides the wave function is normalized to unity, the
statistical ensemble ${\cal S}_{{\rm q}}={\cal E}_{{\rm p}}\left[ {\cal S}_{%
{\rm st}}\right] $ turns to a statistical average system $\left\langle {\cal %
S}_{{\rm st}}\right\rangle $.

Quantum mechanics is a special case of the stochastic systems dynamics.
Quantum systems ${\cal S}_{{\rm q}}$ are special well investigated kind of
stochastic systems. Discussing measurement in non-deterministic system we
shall keep in mind mainly theory of quantum measurements.

Mathematical side of the quantum measurement theory is developed very well 
\cite{N32,M98,M00}. Unfortunately, practically all authors ignore the fact
that the quantum mechanics is a statistical conception, which contains two
sorts of objects: individual physical systems ${\cal S}$, and statistical
average systems $\left\langle {\cal S}\right\rangle $. Mathematical
formalism of quantum mechanics deals only with $\left\langle {\cal S}%
\right\rangle $, describing them in terms of wave functions. All predictions
of quantum mechanics are produced only for statistical average systems $%
\left\langle {\cal S}\right\rangle $, but not for individual systems ${\cal S%
}$.

In the previous section we have seen that the wave function describes the
state of the statistical average system $\left\langle {\cal S}\right\rangle $
, or statistical ensemble ${\cal E}\left[ {\cal S}\right] $, but not the
state of individual system ${\cal S}$. Producing a single measurement, we
deal mainly with an individual system ${\cal S}$.

In fact, in quantum mechanics there are at least two different sorts of
measurements: single measurement ($S$-measurement), i.e. measurement under
individual system ${\cal S}$ and mass measurement ($M$-measurement), i.e.
measurement under many single systems, constituting the statistical ensemble 
${\cal E}\left[ {\cal S}\right] $, or the statistical average system $%
\left\langle {\cal S}\right\rangle $. Properties of $S$-measurement and of $%
M $-measurement quite different and confusion of them leads to
misunderstandings.

The $S$-measurement of the quantity ${\cal R}$ is produced in a single
system $S$ by means of some measuring device ${\cal M}$. $S$-measurement
leads to a definite result $R^{\prime }$ and does not change the wave
function $\psi $, describing the state of $\left\langle S\right\rangle $. As
to the state $X$ of the stochastic system $S$, it is random and is not
considered in quantum mechanics, which describes only reproducible (regular)
features of stochastic systems.

The $M$-measurement of the quantity ${\cal R}$ is a set of $N$ ($%
N\rightarrow \infty $) $S$-measurements, produced in single systems $S$,
constituting the system $\left\langle S\right\rangle $. Result of $M$
-measurement is some distribution $F\left( R\right) $ of values $R^{\prime }$%
, because different $S$-measure\-ments, constituting the $M$-measurement,
give, in general, different values $R^{\prime }$ of the quantity ${\cal R}$.
After $M$-measurement the pure state $\psi $ of $\left\langle S\right\rangle 
$ turns, in general, to a mixed state, described by the matrix density $\rho
=\left| \psi _{R^{\prime }}\right\rangle \left| \left\langle \psi |\psi
_{R}\right\rangle \right| ^{2}\left\langle \psi _{R^{\prime }}\right| $,
where $\left| \psi _{R^{\prime }}\right\rangle $ is the eigenfunction of the
operator $R$ for the quantity ${\cal R}$, corresponding the eigenvalue $%
R^{\prime }$.

According to quantum principles the action of a measuring device ${\cal M}$
on the measured statistical average system $\left\langle {\cal S}%
\right\rangle $ may only change the Hamiltonian $H$, describing evolution of
statistical average system $\left\langle {\cal S}\right\rangle $. However,
no matter what this change of Hamiltonian may be, the state evolution
happens to be a such one that the pure state $\psi $ of statistical average
system $\left\langle {\cal S}\right\rangle $ remains to be pure.
Nevertheless, action of the measuring device ${\cal M}$ on $\left\langle 
{\cal S}\right\rangle $ at the state $\psi $ leads to a passage of $%
\left\langle {\cal S}\right\rangle $ into a mixed state $\rho $.
Qualitatively it is explained by that the measuring device ${\cal M}$
transforms the total Hamiltonian $H$ into several different Hamiltonians $%
H_{1}$, $H_{2},\ldots $, depending on the state $\varphi _{k}$, $%
k=1,2,\ldots $ of measuring device ${\cal M}$.

Let us consider those systems ${\cal S}$ of the statistical ensemble ${\cal E%
}\left[ N,{\cal S}\right] $, which gave the value $R_{1}$ at the measurement
of the quantity ${\cal R}$. They form subensemble ${\cal E}_{1}\left[ N_{1},%
{\cal S}\right] $ of the statistical ensemble ${\cal E}\left[ N,{\cal S}%
\right] $. At the fixed value $R_{1}$ of the measured quantity ${\cal R}$
the measuring device ${\cal M}$ is found at the state $\varphi _{1}$. Then
subensemble ${\cal E}_{1}\left[ N_{1},{\cal S}\right] $ of systems ${\cal S}$
, which gave the result $R_{1}$ at the measurement, evolves with the
Hamiltonian $H_{1}$. The subensemble ${\cal E}_{2}\left[ N_{2},{\cal S}%
\right] $ of systems ${\cal S}$, which gave the result $R_{2}$ at the
measurement, evolves with the Hamiltonian $H_{2}$, because the measuring
device ${\cal M}$ is found now in another state $\varphi _{2}$. Each value $%
R_{k}$ of the measured quantity ${\cal R}$ associates with an evolution of
the subensemble ${\cal E}_{k}\left[ N_{k},{\cal S}\right] $ of the
statistical ensemble ${\cal E}\left[ N,{\cal S}\right] $ with the
Hamiltonian $H_{k}$.

The number $N_{k}$ of systems in the corresponding subensemble ${\cal E}_{k} %
\left[ N_{k},{\cal S}\right] $ is proportional to the probability of the
measured value $R_{k}$ of the measured quantity ${\cal R}$. Evolution of
different subensembles ${\cal E}_{k}\left[ N_{k},{\cal S}\right] $ is
different. As a corollary one cannot speak about one wave function,
describing the state of the whole ensemble ${\cal E}\left[ N,{\cal S}\right] 
$. One should speak about states of subensembles ${\cal E}_{k}\left[ N_{k}, 
{\cal S}\right] $, constituting the statistical ensemble ${\cal E}\left[ N, 
{\cal S}\right] $. Each of subensembles ${\cal E}_{k}\left[ N_{k},{\cal S} %
\right] $ is described by the wave function $\psi _{k},$ And the whole
ensemble ${\cal E}\left[ N,{\cal S}\right] $ is described by the matrix
density, i.e. by a set of wave functions $\psi _{k}$, taken with statistical
weights $\left| \left\langle \psi |\psi _{k}\right\rangle \right| ^{2}$.

In general, any $M$-measurement may be conceived as an abstract single
procedure, produced on the statistical average system $\left\langle {\cal S}
\right\rangle $. Action of $M$-measurement on statistical average system $%
\left\langle {\cal S}\right\rangle $ is described formally by the rule of
von Neumann \cite{N32}. 
\begin{equation}
\psi \rightarrow \left| \psi _{R}\right\rangle \left| \left\langle \psi
|\psi _{R}\right\rangle \right| ^{2}\left\langle \psi _{R}\right|
\label{c3.1}
\end{equation}
Influence (\ref{c3.1}) of the $M$-measurement on the state $\psi $ $%
\left\langle {\cal S}\right\rangle $ is known as a reduction of the wave
function. For the reduction process it is important that result $R_{k}$ of
measurement of the quantity ${\cal R}$ be fixed, i.e. that the measuring
device ${\cal M}$ be found at the state $\varphi _{k}$, because only in this
case one can speak on a definite Hamiltonian $H_{k}$, which determines
evolution of ${\cal E}_{k}\left[ N_{k},{\cal S}\right] $. It is this
procedure of measurement ($M$-measurement) that is considered in most of
papers \cite{N32,M98}.

Is it possible to speak on derivation of a definite value $R_{k}$ at $M$%
-measurement of the quantity ${\cal R}$ for statistical average system $%
\left\langle {\cal S}\right\rangle $? It is possible, but it is a new kind
of measurement, so-called selective $M$-measurement, or $SM$-measurement. $%
SM $-measurement is $M$-measurement, accompanied by a selection of only
those systems ${\cal S}$, for which a single $S$-measurement gives the same
measurement result $R_{k}$. It is of no importance, who or what carries out
this selection. This may be device, human being, or environment. The
selection is introduced directly in the definition of the $SM$-measurement,
and it is to be made by anybody. The process of selection of individual
systems ${\cal S}$ may be interpreted as a statistical influence of the
measuring device ${\cal M}$ on the statistical average system $\left\langle 
{\cal S}\right\rangle $.

The statistical influence in itself is not a force interaction. It is an
influence of the measuring device ${\cal M}$, leading to a selection of some
and discrimination of other elements of the statistical ensemble ${\cal E}%
\left[ \infty ,{\cal S}\right] $. In general, one may say on statistical
influence of the measuring device ${\cal M}$ on the state of the measured
statistical average system $\left\langle {\cal S}\right\rangle $. The state
of the system $\left\langle {\cal S}\right\rangle $ is determined by the
distribution $d\left( X\right) $ of quantities $X$, describing the state of
the system ${\cal {S}}$. Statistical influence of the measuring device $%
{\cal M}$ on $\left\langle {\cal S}\right\rangle $ leads to a change of this
distribution.

Properties of different kinds of measurement of the quantity ${\cal R}$ are
presented in the table 
\[
\begin{array}{|c|c|c|c|c|c|}
\hline
&  &  &  &  &  \\ 
\begin{array}{c}
\text{{\it type of}} \\ 
\text{{\it measur.}}
\end{array}
& 
\begin{array}{c}
\text{{\it object of }} \\ 
\text{{\it measur.}}
\end{array}
& 
\begin{array}{c}
\text{{\it state }} \\ 
\text{{\it before}}
\end{array}
& 
\begin{array}{c}
\text{{\it state after}} \\ 
\text{{\it \ measurement}}
\end{array}
& 
\begin{array}{c}
\text{{\it result }} \\ 
\text{{\it of meas.}}
\end{array}
& 
\begin{array}{c}
\text{{\it influence }} \\ 
\text{{\it on the state }}
\end{array}
\\ 
&  &  &  &  &  \\ \hline
\text{S} & S & \psi & \psi & R^{\prime } & \text{no} \\ \hline
\text{M} & \left\langle S\right\rangle & \psi & \left| \psi
_{R}\right\rangle \left| \left\langle \psi |\psi _{R}\right\rangle \right|
^{2}\left\langle \psi _{R}\right| & f\left( R^{\prime }\right) & \text{%
reduction} \\ \hline
\text{SM} & \left\langle S\right\rangle & \psi & \psi _{R^{\prime }} & 
R^{\prime } & 
\begin{array}{c}
\text{change of } \\ 
\text{pure state.}
\end{array}
\\ \hline
\end{array}
\]

Note that the quantum mechanics makes predictions, concerning only $M$%
-measu\-rements. Any quantum mechanical prediction can be tested only by
means of $M$-measurement. There is no quantum mechanical predictions that
could be tested by means of one single measurement ($S$-measurement). Result
of individual measurement ($S$-measurement) of the quantity ${\cal R}$ can
be predicted never. Using quantum mechanical technique, one can predict only
probability of the fact that the result $R^{\prime }$ will be obtained at
the $S$-measurement of the quantity ${\cal R}$. But a prediction of the
result probability does not mean a prediction of the result in itself.

To understand what means the prediction of the result probability from the
measurement viewpoint, let us consider the following situation. Let a
calculation on the basis of quantum mechanics technique gives, that the
probability of obtaining the result $R^{\prime }$ at the measurement of the
quantity ${\cal R}$ in the system ${\cal S}_{{\rm q}}$ at the state $|\psi
\rangle $ is equal to 1/2. How can one test that this probability is equal
to 1/2, but not, for instance, to 3/4 or 1/4? It is clear that such a test
is impossible at one single measurement of the quantity ${\cal R}$. To test
the prediction, it is necessary to carry out $N$, ($N\rightarrow \infty $)
individual measurements, and the part of measurements, where the value $%
R^{\prime }$ of the quantity ${\cal R}$ is obtained, gives the value of
probability. It is valid even in the case, when predicted probability is
equal to unity. In this case for the test of the prediction an individual
experiment is also insufficient. To test the prediction, one needs to carry
out a set of many individual measurements of the quantity ${\cal R}$, and
the prediction is valid, provided the value $R^{\prime }$ is obtained in all
cases.

At this point the quantum mechanics distinguishes from the classical
mechanics, where results of repeated measurements coincide always. The
classical mechanics accepts that two different individual measurements,
produced on the systems prepared in the same way give similar results. The
classical mechanics supposes that it is possible one not to test this
circumstance, and it predicts the value $R^{\prime }$ of the measured
quantity ${\cal R}$ (probability of the value $R^{\prime }$ is accepted to
be equal to 1 and is not tested). The deterministic foundation of classical
mechanics is manifested in this assumption.

Describing stochastic systems, the quantum mechanics admits that two
different individual measurements, produced on the systems prepared in the
same way, may give different results, and quantum mechanics predicts only
probability of the value $R^{\prime }$ of the measured quantity ${\cal R}$
(but not the value $R^{\prime }$ itself). At the test only value of the
probability is verified (it is tested, even if this value is equal to 1 or
to 0). For such a test one needs a $M$-measurement. In other words, two
predictions: (1) ''measurement of the quantity ${\cal R}$ must give the
value $R^{\prime }$'' and (2) ''measurement of the quantity ${\cal R}$ must
give the value $R^{\prime }$ with the probability 1'' are two different
predictions, tested by measurements of different kinds. Predictions of the
first type can be made only by classical mechanics, which deals {\it only}
with deterministic physical systems. If some of physical systems are
stochastic, then even deterministic physical system is to be considered as a
special case of (stochastic) physical system. Then predictions of the first
type are impossible. The quantum mechanics deals with stochastic systems.
One can make only prediction of the second type, {\it even in the case, when
the physical system evolves as a deterministic one}. This fact is connected
with the above considered statement that in the dynamics of the stochastic
systems the main object of dynamics is $\left\langle {\cal S}\right\rangle $%
, and the evolution of deterministic system ${\cal S}_{{\rm d}}$ is
described as $\left\langle {\cal S}_{{\rm d}}\right\rangle $.

All this means that the quantum mechanical technique and its predictions
deal only with mass measurements ($M$-measurement) and have nothing to do
with individual measurements. In general, appearance of the term
''probability'' in all predictions of quantum mechanics is connected with
the fact that the quantum mechanical technique deals only with distributions 
$d\left( X\right) $ of quantities $X$, which are reproduced at repeated
measurements, but not with the quantities $X$ themselves, whose values are
random and irreproducible. It means that the quantum mechanics technique
deals with statistical average objects $\langle {\cal S}\rangle $ (or with
statistical ensembles of single systems).

Essentially the quantum mechanics investigates dynamic characteristics of $%
\langle {\cal S}\rangle $ and claims them as mean characteristics of ${\cal S%
}$. Sometimes quantum mechanics makes bids for predictions of some
distributions, which appear to be formal and cannot be tested by experiment.
For instance, distribution $|\langle p|\psi \rangle |^{2}$ over the electron
momenta at the pure state, described by the wave function $|\psi \rangle $,
has a formal character, because it cannot be tested experimentally. Here $%
\left\vert p\right\rangle $ means eigenfunction of the momentum operator $%
p=-i\hbar \nabla $. Of course, one can obtain the distribution $|\langle
p|\psi \rangle |^{2}$ experimentally. It is sufficient to drop a flux of
electrons onto diffraction grating and to investigate the obtained
diffraction picture. But the quantum mechanics technique supposes, that
obtained in such a way distribution over momenta must be attributed to some
state (some wave function $|\psi \rangle $). Derivation of the distribution $%
|\langle p|\psi \rangle |^{2}$ needs a long time. This time is the longer
the more exact distribution is to be obtained. In this time the wave
function changes, and it is not clear to what wave function the obtained
distribution over momenta should be attributed. Thus, it is possible to
obtain distribution $|\langle p|\psi \rangle |^{2}$ experimentally, but it
cannot be attributed to any definite state, and it cannot be a test of the
quantum mechanics prediction. The performed analysis \cite{R77} shows that
the momentum distribution, obtained experimentally, cannot be attributed to
any state (wave function).

We have seen in the first section that the statistical average system does
not describe distribution over energies or over momenta. Then how should the
experimentally obtained distribution $|\langle p|\psi \rangle |^{2}$ over
momenta be interpreted?\ The answer is as follows. The distribution $%
|\langle p|\psi \rangle |^{2}$ is not a distribution over momenta. It is a
distribution over {\it mean momenta}. Let us explain what is the mean
momentum. Let the region $V$, occupied by the electron beam, be separated
into regions $V_{1}$,$V_{2}$,...,$V_{n}$,... Each of regions $V_{k}$ is
small, but its linear size $l_{k}\gg \lambda _{c}$, where $\lambda _{c}$ is
the Compton wavelength. Let us calculate the mean value $\left\langle {\bf p}%
_{k}\right\rangle $ of the momentum ${\bf p}$ of an electron inside the
region $V_{k}$. Such a quantity can be calculated, because the momentum
density is an additive quantity, and the mean value $\left\langle {\bf p}%
_{k}\right\rangle $ can be calculated for any region. We refer to \cite{R77}
for substantiation of this statement. Here we only illustrate the difference
between the distribution over momenta and distribution over mean momenta in
a simple example.

Let us consider flow of a gas in a region $V$. Let us separate the region $V$
into small but macroscopic regions $V_{1}$,$V_{2}$,...,$V_{n}$,...Let us
calculate the mean value $\left\langle {\bf p}_{k}\right\rangle $ of the
molecule momentum ${\bf p}$ inside the region $V_{k}$. The set of all $%
\left\langle {\bf p}_{k}\right\rangle $, taken at some time moment gives
distribution of the gas molecules over mean momenta. It is clear that the
real distribution of gas molecules over velocities (Maxwell distribution)
has nothing to do with distribution of gas particles over macroscopic mean
velocities $\left\langle {\bf v}_{k}\right\rangle =\left\langle {\bf p}%
_{k}\right\rangle /m$. Let us imagine that we measure the molecule positions 
${\bf x}$ and ${\bf x}^{\prime }$, separated by the time interval $\Delta t$
and determine the molecule velocity by means of relation ${\bf v}=({\bf x}-%
{\bf x}^{\prime })/\Delta t$. If $\Delta t$ much less than, the mean time $%
\tau $ between the molecule collisions, the obtained distribution over
velocities is the Maxwell distribution. Vice versa, if $\Delta t\gg \tau $,
the obtained velocity is the macroscopic mean gas velocity, and the obtained
distribution is a distribution over mean gas velocities in different regions 
$V_{k}$.

Something like that we obtain in the case of the electron beam. A use of the
diffraction grating is equivalent to measurement of the electron positions
separated by the time interval $\Delta t\rightarrow \infty $. It leads to
distribution over mean momenta. If $\Delta t$ is small enough, and the
distance $\Delta {\bf x}={\bf x-x}^{\prime }$ between the subsequent
electron positions ${\bf x}$ and ${\bf x}^{\prime }$ is small, then the
electron momentum increases according to the indeterminacy principle, and we
begin to measure the stochastic component of the electron motion. At very
small $\Delta t$ we measure only stochastic component of motion, which
depends only on $\Delta t$ and does not depend practically on the wave
function of the electron beam.

Let us consider now well known paradox of ''Schr\"odinger cat''. At first,
let us present it in the conventional manner. There is a cat in a closed
chamber. The cat's life is determined by the state of a radioactive atom,
placed in this chamber. While the atom is not decayed, the cat is alive. As
soon as the atom decays, the cat becomes dead. The state of the atom is a
linear superposition of states of the indecomposed atom and decomposed atom.
Respectively the state of the cat is a linear superposition of the alive cat
and of the dead one. Paradox consists in the simultaneous existence of the
dead cat and of the alive cat. If one opens the chamber and observes the
cat, the cat passes instantly from the state, where the cat is neither dead,
nor alive, to the definite state, where the cat is either dead, or alive.

This paradox is a result of simple misunderstanding, when one identifies two
different object: the real individual Cat and an abstract statistical
average $\langle \mbox{cat}\rangle $. The wave function describes the state
of the abstract statistical average $\langle \mbox{atom}\rangle ,$ and the
state of the abstract statistical average $\langle \mbox{atom}\rangle $
determines the state of the statistical average $\langle \mbox{cat}\rangle $
. The wave function has nothing to do with the real Cat in the chamber. The
statistical average $\langle \mbox{cat}\rangle $ bears on the real Cat the
same relation as the statistical average inhabitant of Moscow bears on a
real Ivan Sidorov, living somewhere in Leninsky avenue. If we produce $S$%
-observation, i.e. we open one definite chamber and found there an alive
cat, we have no reasons for the statement, that opening the chamber, we
change the state of $\langle \mbox{cat}\rangle $. No paradox appears, when
the chamber is open. If we produce $M$-observation, i.e. if we consider $N$,
($N\gg 1$) chambers with cats, then opening them all simultaneously, we do
not discover a definite result. In some chambers one discovers alive cats,
in other ones the cats are dead. In this case observation of the state of
the abstract statistical average $\langle ${\rm cat}$\rangle $ leads to a
change of the state in the sense that the state turns from pure to mixed.
But there is no paradox.

Finally, if we carry out a selective mass measurement ($SM$-measurement),
i.e. one opens simultaneously $N$, ($N\gg 1$) chambers and chooses those of
them, where there are alive cats, then, on one hand, one obtains a definite
result (alive cats), but on the other hand, there is many alive cats, and
they form a statistical ensemble, whose state is described by a certain
definite wave function. But any paradox does not appears, as far as the
reason of a change of the wave function of statistical average $\langle $%
{\rm cat}$\rangle $ is evident. It is a selection of alive cats from the
total set, consisting of alive and dead cats.

The Schr\"{o}dinger cat paradox is a special case of the paradox, which is
discerned sometimes in the wave function reduction, appearing as a result of
a measurement. To avoid this paradox, it is sufficient to follow a simple
logic rule, which asserts: ''One may not use the same term for notation of
different objects. (But it is admissible to use several different terms for
notation of the same object.)''. Usually in quantum mechanics this rule is
violated. One uses the same term for the individual object and for the
statistical average object. As a corollary one uses the same term for two
different measurement processes.

Note that the formal representation of $M$-measurement as a single act of
influence on the statistical average system $\left\langle {\cal S}%
\right\rangle $ makes a large contribution into identification of $S$%
-measurement with $M$-measurement. This representation removes distinction
between the $S$-measurement and the $M$-measurement and gives the impression
that $S$-measurement and the $M$-measurement are identical procedures. After
such an identification the measurement process acquires contradictory
properties. On one hand, such a measurement leads to a definite result ($S$%
-measurement), on the other hand, its result is a distribution of the
measured quantity ($M$-measurement). On one hand, the measurement ($S$%
-measurement) is a single act, produced on ${\cal S}$, and the measurement
does not bear on the wave function, and on the statistical average system $%
\left\langle {\cal S}\right\rangle $. On the other hand, the measurement ($M$%
-measurement) leads to a reduction of the state of the system $\left\langle 
{\cal S}\right\rangle $.

In general, violation of the formal logic rule cited above is an origin for
appearance of paradoxes and contradictions. Some exotic interpretation of
quantum mechanics were removed by themselves (for instance, \cite{E57,DG73}%
), provided one follows this logic rule and distinguish between individual
object and statistically average one.

It should note that sometimes it is difficult to determine whether some
property is a property of an individual object or that of a statistical
average one. For instance, whether the half-integer spin of an electron is a
property of a single electron or a property of statistical average electron
(statistical ensemble). At the conventional approach, when one does not
distinguish between an individual electron and a statistical average
electron, such a problem does not exist at all. At a more careful approach
we are forced to state that properties of an individual electron have been
investigated very slightly. In most of cases we deal with measurements,
produced on many electrons, as far as only such experiments are
reproducible. As a corollary many conclusions on properties of a single
electron are unreliable.

For instance, it is a common practice to consider the Stern -- Gerlach
experiment (when the electron beam, passing the region with inhomogeneous
magnetic field, splits into two beams) to be an evidence of the statement
that each single electron has a definite spin and corresponding projection
of magnetic moment onto the magnetic field direction. In reality, the Stern
-- Gerlach experiment shows only that the Hamiltonian, describing a motion
of statistical average electron, has two discrete eigenstates, labelled by
the magnetic quantum number and distinguishing by their energy. The
question, whether discreteness of the magnetic quantum number is connected
with a discreteness in properties of a single electron, remains open,
because a discreteness of energy levels is not connected directly with the
discrete character of interaction. For instance, the energy levels of a
spinless charged particle in the Coulomb electric field are discrete,
although there is no discreteness in the properties of a single particle. It
is doubtless that the beam split is connected with 'the electron magnetic
moment', because it is proportional to the magnetic field gradient. But the
question remains open, whether the 'electron magnetic moment' is a property
of the individual electron ${\cal S}$ or a collective property of the
statistical average electron $\left\langle {\cal S}\right\rangle $ (see for
details \cite{R95}). Note that at identification of $\left\langle {\cal S}%
\right\rangle $ and ${\cal S}$ the problem does not appear.

Sometimes one considers, that the spectrum of electromagnetic radiation,
emanated by an excited atom is a property of individual (but not statistical
average) atom. The argument is adduced that the modern technique admits one
to confine a single atom in a trap. Then one can investigate its energy
levels and spectrum of radiation. But one overlooks that the spectrum of the
atom radiation cannot be measured as a result of a single measurement ($S$
-measurement). The atom radiation spectrum is obtained as a result of many
measurements of radiation of the same atom, whose state is prepared by the
same way, i.e. that essentially one measures radiation of a statistical
average atom. To carry out such a $M$-measurement, it is of no importance,
whether one produces one measurement with many similarly prepared atoms, or
one produces many measurements with a single atom, preparing its state many
times by the same way. In both cases one deals with the statistical average
atom.

\section{Concluding remarks}

Thus, discrete physical system ${\cal S}$ has a regular component of its
evolution and a stochastic one. The regular component is described
explicitly by statistical average system $\left\langle {\cal S}\right\rangle 
$, which is a continuous dynamic system. The stochastic component of system $%
{\cal S}$ evolution is also described by $\left\langle {\cal S}\right\rangle 
$. This description is implicit. The action of the system $\left\langle 
{\cal S}\right\rangle $ is reduced to the form of the action for dynamic
system ${\cal S}_{{\rm red}}\left[ {\cal S}_{{\rm d}}\right] $, which is a
set of interacting identical discrete dynamic systems ${\cal S}_{{\rm d}}$.
The term in the action, responsible for interaction of ${\cal S}_{{\rm d}}$,
describes implicitly the character of stochastic component of the system $%
{\cal S}$ evolution. The system ${\cal S}$ has finite number of the freedom
degrees, whereas the system $\left\langle {\cal S}\right\rangle $,
describing its dynamics, has infinite number of the freedom degrees. From
informative viewpoint the system $\left\langle {\cal S}\right\rangle $ is
more complicated, than the system ${\cal S}$, but we are forced to
investigate $\left\langle {\cal S}\right\rangle $, if we want to study the
stochastic system dynamics.

\end{document}